\documentclass[preprint,aps,showpacs,nofootinbib,preprintnumbers,amsmath,amssymb,superscriptaddress]{revtex4-1}
\usepackage{amssymb}
\usepackage{epsfig}
\usepackage{latexsym}
\usepackage{graphicx}% Include figure files
\usepackage{subfigure}
\usepackage{dcolumn}% Align table columns on decimal point
\usepackage{bm}% bold math
\usepackage[usenames ,dvipsnames]{xcolor}
\usepackage[utf8]{inputenc}
\usepackage{slashed}
\usepackage{cleveref}
\usepackage{enumitem}

\begin{document}
\title{$W$-boson Mass Anomaly from High-Dimensional Scalar Multiplets}
\author{Jiajun~Wu\footnote{wujiajun@itp.ac.cn}}
\affiliation{School of Fundamental Physics and Mathematical Sciences, Hangzhou Institute for Advanced Study, UCAS, Hangzhou 310024, China}
\affiliation{University of Chinese Academy of Sciences (UCAS), Beijing 100049, China}
\affiliation{International Centre for Theoretical Physics Asia-Pacific, Beijing/Hangzhou, China}
\author{Chao-Qiang~Geng\footnote{geng@phys.nthu.edu.tw}}
\affiliation{School of Fundamental Physics and Mathematical Sciences, Hangzhou Institute for Advanced Study, UCAS, Hangzhou 310024, China}
\affiliation{University of Chinese Academy of Sciences (UCAS), Beijing 100049, China}
\affiliation{International Centre for Theoretical Physics Asia-Pacific, Beijing/Hangzhou, China}
\author{Da~Huang\footnote{dahuang@bao.ac.cn}}
\affiliation{School of Fundamental Physics and Mathematical Sciences, Hangzhou Institute for Advanced Study, UCAS, Hangzhou 310024, China}
\affiliation{International Centre for Theoretical Physics Asia-Pacific, Beijing/Hangzhou, China}
\affiliation{National Astronomical Observatories, Chinese Academy of Sciences, Beijing, 100012, China}

\date{\today}
\begin{abstract}
In light of the recently discovered $W$-boson mass anomaly by the CDF Collaboration, we discuss two distinct mechanisms that could possibly explain this anomaly through the introduction of high-dimensional $SU(2)_L$ scalar multiplets. The first mechanism is the tree-level $W$-boson mass correction induced by the vacuum expectation values of one or more $SU(2)_L$ scalar multiplets with odd dimensions of $n\geq 3$ and zero hypercharge of $Y=0$ in order to avoid the strong constraint from  measurements of the $Z$-boson mass. The second mechanism is to consider the one-loop level $W$-boson mass correction from a complex multiplet. In particular, we focus on the case with an additional scalar octuplet with $Y=7/2$. As a result, we find that both mechanisms can explain the $W$-boson mass anomaly without violating any other theoretical or experimental constraints.
\end{abstract}

\maketitle

\section{Introduction}
Recently, the CDF-\uppercase\expandafter{\romannumeral 2} Collaboration has reported the most precise measurement of the $W$-boson mass, showing that the observed $W$-boson mass of $m^{\rm CDF-\uppercase\expandafter{\romannumeral2}}_W = 80433.5\pm9.4$~MeV~\cite{CDF:2022hxs} deviates the latest Standard Model (SM) prediction of $m^{\rm SM}_W= 80357\pm 6$~MeV~\cite{ParticleDataGroup:2022pth}. The significance of this anomaly is more than $7\sigma$, which indicates new physics (NP) beyond the SM (BSM). Therefore, it is a natural and pressing concern to introduce NP models to account for this anomalous $W$-boson mass.

Among a multitude of BSM scenarios, extending the SM Higgs sector by incorporating extra $SU(2)_L$ multiplets~\cite{Sakurai:2022hwh,Peli:2022ybi,Dcruz:2022dao,Asai:2022uix,Fan:2022dck,Lu:2022bgw,Song:2022xts,Bahl:2022xzi,Babu:2022pdn,Heo:2022dey,Ahn:2022xax,Ghorbani:2022vtv,Lee:2022gyf,Abouabid:2022lpg,Benbrik:2022dja,Botella:2022rte,Kim:2022hvh,Kim:2022xuo,Appelquist:2022qgl,Benincasa:2022elt,Arhrib:2022inj,Han:2022juu,Abdallah:2022shy,deGiorgi:2023wjh,Abouabid:2023mbu,Barrie:2022cub,Cheng:2022jyi,Du:2022brr,FileviezPerez:2022lxp,Kanemura:2022ahw,Mondal:2022xdy,Borah:2022obi,Addazi:2022fbj,Heeck:2022fvl,Chen:2022ocr,Evans:2022dgq,Ghosh:2022zqs,Ma:2022emu,Bahl:2022gqg,Penedo:2022gej,Cheng:2022hbo,Butterworth:2022dkt,Wu:2022uwk,Song:2022jns,Crivellin:2023gtf,Ellis:2023zim,Shimizu:2023rvi,Senjanovic:2022zwy,Ashanujjaman:2023etj} is a promising avenue, since the modification of the scalar sector is intricately connected to the underlying mechanism for the electroweak (EW) gauge symmetry breaking and the related hierarchy problem, which can be studied through the measurement of EW oblique parameters~\cite{Peskin:1990zt,Peskin:1991sw,Maksymyk:1993zm,Burgess:1993mg,Lavoura:1993nq,Albergaria:2021dmq}. Furthermore, the added scalar multiplet has the potential to resolve many puzzles in the SM, such as the nature of dark matter (DM)~\cite{Cirelli:2005uq,Cirelli:2009uv,Guo:2010hq,Barbieri:2006dq,LopezHonorez:2010eeh,Gonderinger:2012rd}, the origin of the matter-antimatter asymmetry~\cite{Cline:2012hg,Grzadkowski:2018nbc,Cline:2021iff,Morrissey:2012db}, and the characteristics of the EW phase transition as well as its related stochastic gravitational wave signals~\cite{Chowdhury:2011ga,Hashino:2018zsi,Chiang:2017nmu,Kannike:2019wsn,Chiang:2020yym,Chiang:2019oms,Cai:2017tmh,Chao:2017vrq,Ellis:2018mja,Alves:2018jsw,Zhou:2018zli,Bian:2019kmg,Ghosh:2020ipy,Zhou:2020irf,Lu:2022zpn,Zhou:2022mlz,Cai:2022bcf,Hashino:2018wee}. Therefore, comprehending the structure of the scalar sector could lead to a more profound understanding of the big picture of the SM and the physics beyond it. 
Extensive studies have been carried out in the literature to explain the CDF-\uppercase\expandafter{\romannumeral 2} $W$-boson mass anomaly with low-dimensional scalar multiplets, which include a scalar singlet~\cite{Sakurai:2022hwh,Peli:2022ybi,Dcruz:2022dao,Asai:2022uix}, a second Higgs doublet~\cite{Fan:2022dck,Lu:2022bgw,Song:2022xts,Bahl:2022xzi,Babu:2022pdn,Heo:2022dey,Ahn:2022xax,Ghorbani:2022vtv,Lee:2022gyf,Abouabid:2022lpg,Benbrik:2022dja,Botella:2022rte,Kim:2022hvh,Kim:2022xuo,Appelquist:2022qgl,Benincasa:2022elt,Arhrib:2022inj,Han:2022juu,Abdallah:2022shy,deGiorgi:2023wjh,Abouabid:2023mbu}, and a scalar triplet~\cite{Barrie:2022cub,Cheng:2022jyi,Du:2022brr,FileviezPerez:2022lxp,Kanemura:2022ahw,Mondal:2022xdy,Borah:2022obi,Addazi:2022fbj,Heeck:2022fvl,Chen:2022ocr,Evans:2022dgq,Ghosh:2022zqs,Ma:2022emu,Bahl:2022gqg,Penedo:2022gej,Cheng:2022hbo,Butterworth:2022dkt,Song:2022jns,Crivellin:2023gtf,Ellis:2023zim,Shimizu:2023rvi,Senjanovic:2022zwy,Ashanujjaman:2023etj}. More recently, we have explored scalar multiplet scenario up to a maximum of a septuplet~\cite{Wu:2022uwk}. 
In this letter, we aim to explain the $W$-boson mass anomaly with higher dimensional multiplets, at both tree and one-loop levels. In particular, for the one-loop solution, we shall focus on the case of a scalar octuplet with $Y=7/2$ and zero vacuum expectation value (VEV), which is the highest dimension for a complex scalar allowed by the perturbative unitarity constraint~\cite{Hally:2012pu}.   

This paper is organized as follows: In Sec.~\ref{s2}, we investigate the possibility of the tree-level explanation of the CDF-\uppercase\expandafter{\romannumeral 2} $W$-boson mass excess with a high-dimensional multiplet. In Sec.~\ref{s3}, we present a detailed phenomenological study of a scalar octuplet with $Y=7/2$, which is of physical interest to explain the CDF-\uppercase\expandafter{\romannumeral 2} $W$-boson mass anomaly at the one-loop level. Finally, we conclude in Sec.~\ref{s4}.

\section{Tree-Level Explanation of $W$-boson Mass Anomaly}\label{s2}
One may easily explain the $W$-boson mass anomaly by introducing a $SU(2)_L$ scalar multiplet with its VEV of the neutral component inducing an additional mass correction to the $W$-boson mass. However, such an idea is hampered by the fact that, since the $W$ and $Z$-bosons have the common origin from the EW gauge symmetry breaking, the associated $Z$-boson mass should also be corrected, which was strongly constrained by the current experiments~\cite{ParticleDataGroup:2022pth}. In this section, we show that if the added scalar multiplet is in an odd-dimensional presentation of $SU(2)_L$ with zero hypercharge, the above problem can be resolved automatically.

Let us begin by considering a real $SU(2)_L$ multiplet $\xi$ of dimension $n=2k+1$ with $k$ as a positive integer denoting the weak isospin $SU(2)_L$ representation. The hypercharge of $\xi$ is fixed to $Y=0$. When $\xi$ and the SM Higgs doublet obtain their VEVs, $v_\xi$, and $v_H$, from the spontaneous breaking of the EW gauge symmetry, the $W$ and $Z$-boson mass terms can be written as follows
\begin{eqnarray}
	D^\mu H^\dagger D_\mu H + D^\mu \xi^\dagger D_\mu \xi \supset &-&\left(\frac{1}{4} g^2 v_H^2 + \frac{1}{2} k(1+k)v_\xi^2 \right) W_\mu^+ W^{-\,\mu} \nonumber \\ 
	&-& \frac{1}{8} (g^2 + g^{\prime \,2}) v_H^2 Z_\mu Z^\mu\,, 
\end{eqnarray}
where $D_\mu$ denotes the covariant derivatives of the scalar fields $H$ and $\xi$ with $g$ and $g^\prime$ the SM $SU(2)_L$ and $U(1)_Y$ gauge couplings, respectively. 
%Note that in the above formula, we have used the normalization of the $SU(2)_L$ generators as that in Ref.~\cite{Georgi:1999wka}. 
Therefore, the SM gauge boson masses are given by
\begin{eqnarray}
	m_W= \frac{1}{2} g\sqrt{v_H^2+2k(1+k)v_\xi^2}\,,\quad m_Z= \frac{v_H}{2}\sqrt{g^2 + g^{\prime\,2}}\,,
\end{eqnarray} 
indicating that the $Z$-boson mass is not corrected in the presence of the extra multiplet scalar VEV of $v_\xi$. It can be understood by the fact that, for $Y=0$, the couplings of various components in the multiplet with the $Z$-boson are proportional to their electric charges, so that the neutral component and the associated VEV cannot interact with the $Z$ boson. As a consequence, this scalar-multiplet-extended model can avoid the strong constraint from the $Z$-boson mass measurement~\cite{ParticleDataGroup:2022pth}. Further, if we take the CDF-\uppercase\expandafter{\romannumeral 2} value of the $W$-boson mass as the one in our model, then the VEV of the multiplet can be estimated as follows
\begin{eqnarray}\label{mW2sig}
	\frac{(\Delta v)^2}{v_H^2} \equiv \frac{2k(1+k)v_\xi^2}{v_H^2}  =  \left(\frac{m_W^{\rm CDF} }{m_W^{\rm SM}}\right)^2-1 \sim ~ [0.00090, 0.00201]\,,  \quad \mbox{at $2\sigma$ C.L.}\,,
\end{eqnarray}
where the SM Higgs doublet VEV is taken to be $v_H = 246.22$~GeV~\cite{ParticleDataGroup:2022pth}. Moreover, note that the model can be further extended by introducing a series of $SU(2)_L$ scalar multiplets with vanishing hypercharges. In this case, we can still keep the salient feature that the $Z$-boson mass is not modified so that the related constraint is weak.

Another critical constraint on the added scalar multiplet is provided by the constraint of the $\rho$ parameter, which can be related to the oblique parameter $T$ as follows~\cite{Grimus:2007if,ParticleDataGroup:2022pth}
	\begin{eqnarray}\label{RhoT}
		\Delta\rho = \alpha T.
	\end{eqnarray}
	In the light of the current constraint on $T$ given by the updated electroweak global fit in Ref.~\cite{Cheng:2022hbo} which includes the CDF-\uppercase\expandafter{\romannumeral2} $W$-boson mass and fixes $S=U=0$, we have
	\begin{eqnarray}\label{RhoValue}
		\Delta\rho \in [0.00101, 0.00133]\,,\quad \mbox{at $2\sigma$ C.L.}\
	\end{eqnarray}
	Theoretically, our model predicts this important quantity as follows~\cite{ParticleDataGroup:2022pth}
	\begin{eqnarray}
		\rho = 1+ \frac{2k(1+k)v_\xi^2}{v_H^2} \,.
	\end{eqnarray} 
	According to Eq.~(\ref{RhoValue}), we can obtain the following $2\sigma$ region for $(\Delta v)^2/v_H^2$
	\begin{eqnarray}\label{ConsRho}
		\frac{(\Delta v)^2}{v_H^2} \in [0.00101, 0.00133]\,,\quad \mbox{at $2\sigma$ C.L.}\,.
	\end{eqnarray}
	Obviously, the multiplet VEV range in Eq.~(\ref{ConsRho}) allowed by the $\rho$ parameter constraints is not in conflict with the CDF-\uppercase\expandafter{\romannumeral2} $W$-boson mass signal region in Eq.~(\ref{mW2sig}). % are not contradictory to each other, which 
	Consequently, the CDF-\uppercase\expandafter{\romannumeral2} $m_W$ anomaly can be solved in this multiplet scalar extension of the SM at the tree-level, while the data of $\rho$ strongly constrains the allowed parameter space.

\section{One-Loop Level Explanation of the $W$-boson Mass Anomaly}\label{s3}
In this section, we discuss the explanation of the $W$-boson mass anomaly with only the one-loop corrections from the addition of high-dimensional scalar multiplet fields. %at the one-loop level. Furthermore, 
After a general review of the relation between EW oblique parameters and the one-loop $W$-boson mass correction, we shall focus as a concrete example on the case of a scalar octuplet with $Y=7/2$ and a vanishing VEV.

\subsection{Oblique Parameters and the $W$-Boson Mass}\label{ss3.1}
The NP effects in the EW sector are usually encoded by the three oblique parameters, namely $S$, $T$, and $U$~\cite{Peskin:1991sw,Peskin:1990zt}, which are defined at the one-loop level as follows: %These oblique parameters at the one-loop level can be defined as follows:
\begin{eqnarray}
S&\equiv&\frac{4s_{W}^{2}c_{W}^{2}}{\alpha}\left[A_{ZZ}^{\prime}\left(0\right)-\frac{c_{W}^{2}-s_{W}^{2}}{c_{W}s_{W}}A_{Z\gamma}^{\prime}\left(0\right)-A_{\gamma\gamma}^{\prime}\left(0\right)\right]\,, \nonumber\\
T&\equiv&\frac{1}{\alpha m_{Z}^{2}}\left[\frac{A_{WW}\left(0\right)}{c_{W}^{2}}-A_{ZZ}\left(0\right)\right]\,, \nonumber\\
U&\equiv&\frac{4s_{W}^{2}}{\alpha}\left[A_{WW}^{\prime}\left(0\right)-\frac{c_{W}}{s_{W}}A_{Z\gamma}^{\prime}\left(0\right)-A_{\gamma\gamma}^{\prime}\left(0\right)\right]-S\,,
\end{eqnarray}
where the functions $A^{(\prime)}_{VV^\prime} (q^2)$ refer to the vacuum polarizations for EW gauge bosons $V^{(\prime)} = \gamma,\, W,\,Z$. 
Moreover, as shown in Refs.~\cite{Fan:2022dck,Peskin:1991sw}, the general one-loop corrections of non-SM scalars to the $W$-boson mass squared can be expressed in terms of these oblique parameters $S$, $T$, and $U$ as follows
\begin{eqnarray}\label{00}
\Delta m_W^2 = \frac{\alpha c_W^2 m_Z^2}{c_W^2 - s_W^2} \left[ -\frac{S}{2}  + c_W^2 T + \frac{c_W^2 - s_W^2}{4s_W^2} U \right]\,,
\end{eqnarray}
where $s_W$ ($c_W$) are (co)sine of the Weinberg angle. 

However, as demonstrated in Ref.\cite{Asadi:2022xiy}, the parameters $T$ and $S$ are usually generated by dimension-6 operators, while $U$ can only be induced by a dimension-8 operator, resulting in its significant suppression. Therefore, in a NP model augmented by a scalar multiplet, it is expected that the dominant one-loop contribution to the $W$-boson mass arises from $T$ and $S$, and the impact of $U$ can be neglected. Besides, Ref.~\cite{Wu:2022uwk} has explicitly confirmed this expectation by demonstrating that when the masses of extra scalars exceed 300~GeV and their multiplet dimensions are restricted to be smaller than 10, the correction of $U$ to the $W$ mass is at least one order of magnitude smaller than the leading ones from $T$ and $S$. %, which explicitly confirms the previously stated expectation. 
According to Refs.\cite{Wu:2022uwk,Lavoura:1993nq}, the contribution of a scalar multiplet $\xi$ to $T$ can be expressed as:
\begin{equation}\label{T}
T_{\xi}=\frac{1}{4\pi s_{w}^{2}m_{W}^{2}}\sum_{I=-k}^{k-1}N_{I+1}^{2}F\left(m_{\xi_{I}^{Q}}^{2},m_{\xi_{I+1}^{Q}}^{2}\right)\,,
\end{equation}
where $I=k,k-1,k-2,......,-k+1,-k$, $N_{I}=\sqrt{{(k+I)(k-I+1)}/{2}}$, $Q=I+Y$ is referred to the electric charge, and the function $F(A,B)$ is given by
\begin{equation}\label{Ffunc}
F\left(A,B\right)\equiv\begin{cases}
	\frac{A+B}{2}-\frac{AB}{A-B}\ln\frac{A}{B}\,, & A\neq B\,,\\
	0\,, & A=B\,.
\end{cases}
\end{equation}
On the other hand, the correction of $\xi$ to $S$ can be expressed as follows:
\begin{equation}\label{S}
S_{\xi}=-\frac{Y}{3\pi}\sum_{I=-k}^{k}I\ln m_{\xi_{I}^{Q}}^{2}\,,
\end{equation}
where $\xi_{I}^{Q}$ represents components of the scalar multiplet $\xi$ with its electric charge $Q$.

\subsection{Scalar octuplet Explanation of the $W$-boson Mass Anomaly}
According to Ref.~\cite{Hally:2012pu}, the highest dimension of a complex scalar multiplet allowed by the perturbative unitarity constraint is eight. %for the scalar mass larger than 1700~GeV. %octuplet, and nonet for real scalar cases. 
Moreover, Ref.~\cite{Wu:2022uwk} has shown that a real scalar multiplet is unable to generate the mass splitting to produce nonzero values of $T$ and $S$, which is required to explain the $W$-mass anomaly. Consequently, in the following, we will focus on the model by adding a complex octuplet scalar with $Y=7/2$, which has not been discussed so far in the literature.

Note that the potential in this model can be constructed as follows with the Higgs doublet $H$ and the scalar octuplet $\xi$
\begin{equation}\label{potential}
\begin{aligned}
	V\left(H,\xi\right)=&-\mu_{H}^{2}H^{\dagger}H+\lambda_{H}\left(H^{\dagger}H\right)^{2}+\mu_{\xi}^{2}\xi^{\dagger}\xi+\lambda_{1}\left(\xi^{\dagger}\xi\right)^{2} +\lambda_{2}\left(\xi^{\dagger}T_{\Phi}^{a}\xi\right)^{2} \\
	&+\lambda_{3}\left(\xi^{\dagger}\xi\right)\left(H^{\dagger}H\right)
	+\lambda_{4}\left(\xi^{\dagger}T_{\xi}^{a}\xi\right)\left(H^{\dagger}T_{H}^{a}H\right)+\lambda_{5}\left(\xi^{\dagger}T_{\xi}^{a}T_{\xi}^{b}\xi\right)^{2}\,.
\end{aligned}
\end{equation}
Here, we present the most generic interaction terms for a scalar octuplet $\xi$. It is worth noting that the potential given in Eq.(\ref{potential}) exhibits an $Z_2$ symmetry that arises naturally without the need for any additional assumptions. 
Note that the mass splitting can be only generated in Eq.~(\ref{potential}) by the following term:
\begin{eqnarray}\label{O4}
O_{4}=\lambda_{4}\left(\xi^{\dagger}T_{\xi}^{a}\xi\right)\left(H^{\dagger}T_{H}^{a}H\right)\,.
\end{eqnarray}
%the scalar multiplet will give no contribution to the parameters $T$ and $S$, as well as the $W$-boson mass, if there are not mass splittings among components in $\xi$. 
Therefore, we will concentrate on this term in our subsequent phenomenological studies. Additionally, it is worth noting that $|\lambda_4|$ is also constrained by perturbativity with $|\lambda_4| < 4\pi$~\cite{Nebot:2007bc}. Hence, In our further discussions, we take into account this perturbative limit by requiring $|\lambda_4| \leq 10$.

For the model with a scalar octuplet $\xi$ of $Y$=7/2, the phenomenology can be classified into the following two distinct types based on the sign of $\lambda_4$:
\begin{itemize}
\item Type A: when $\lambda_4 >0$, the lightest particle in the octuplet is the most electrically charged one with its mass denoted as $M_C$, so that $M_L = M_C$.
\item Type B: when $\lambda_4 <0$, the lightest particle in the octuplet is the electrically neutral one with its mass denoted as $M_0$, indicating $M_L = M_0$.
\end{itemize}
In light of the mass splittings among scalars from $O_4$, this $Y=7/2$ scalar octuplet has the potential to explain the $W$-mass anomaly with its following nonzero corrections to the parameters $T$ and $S$: %with $U$ held constant at a fixed value of zero. 
%The one-loop corrections of additional nonet in $Y=4$ case to the $W$-boson squared can be expressed in terms of $T$ and $S$, as shown below:
\begin{eqnarray}
\Delta m_W^2 = \frac{\alpha c_W^2 m_Z^2}{c_W^2 - s_W^2} \left[ -\frac{S}{2}  + c_W^2 T \right]\,.
\end{eqnarray}
\begin{figure}[h]
\centerline{\includegraphics[width=0.9\linewidth]{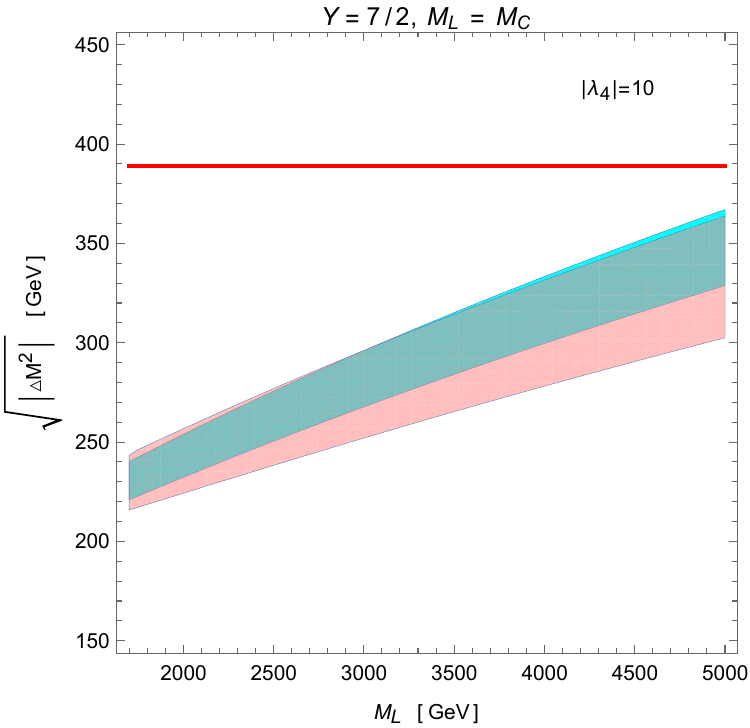}}
\caption{The parameter region in the $M_{L}-\sqrt{|\Delta M^{2}|}$ plane for the Type-A scalar octuplet model, in which the lightest scalar is the most charged one, {i.e.}, $M_{L}=M_{C}$. The horizontal axis denotes the mass of the lightest particle in the octuplet ($M_{L}$), ranging from 1700 GeV to 5000 GeV, while the vertical axis denotes the mass splitting between adjacent components. The pink region in the figure indicates the parameter space allowed by electroweak global fits for $T$ and $S$ at the $2\sigma$ CL when $U=0$, as reported in Ref.~\cite{Asadi:2022xiy}. The solid cyan area corresponds to the parameter space that can explain the measured $W$-boson mass by ${\rm CDF\mbox{-}\uppercase\expandafter{\romannumeral2}}$ within the $2\sigma$ CL.%, and the shaded regions represent the parameter space that satisfies the corresponding requirements. 
		The red solid line represents the mass difference for $|\lambda_{4}|=10$, and the region below this line satisfies the perturbativity condition.}\label{f2}
\end{figure}
\begin{figure}[h]
\centerline{\includegraphics[width=0.9\linewidth]{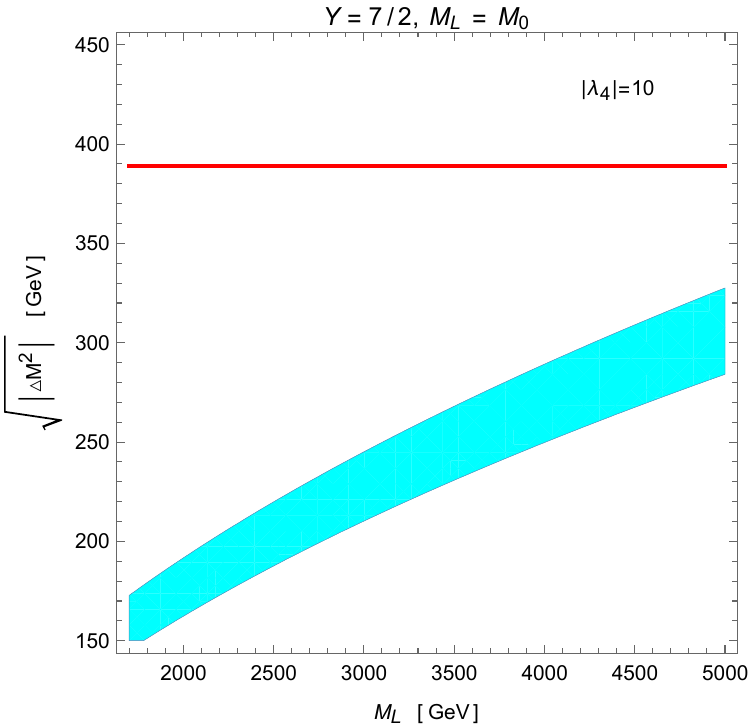}}
\caption{Legend is the same as Fig.~\ref{f2} but for the Type-B scalar octuplet model. \label{f3}}
\end{figure}

Figs.~\ref{f2} and \ref{f3} display the parameter spaces in the $M_L$-$\sqrt{\Delta m^2}$ plane for the Type-A and Type-B scalar octuplet models, respectively. The horizontal axis denotes the mass of the lightest particle in the octuplet ($M_{L}$), ranging from 1700 GeV to 5000 GeV, while the vertical axis denotes the mass splitting between adjacent components. The pink region in the figure indicates the parameter space allowed by EW global fits for $T$ and $S$ at the $2\sigma$ CL when $U=0$, as reported in Ref.~\cite{Asadi:2022xiy}. The cyan area corresponds to the parameter space that can explain the measured $W$-boson mass by ${\rm CDF\mbox{-}\uppercase\expandafter{\romannumeral2}}$ within the $2\sigma$ CL. %and the shaded regions represent the parameter space that satisfies the corresponding requirements. 
The red solid line represents the scalar mass difference corresponding to the perturbative limit $|\lambda_{4}|=10$.%, and the regions below the red line satisfy the perturbativity condition. 
The results show that the Type-A model has a substantial amount of parameter space that can solve the CDF-\uppercase\expandafter{\romannumeral2} $m_W$ anomaly while satisfying the EW global fits and perturbative limits. On the other hand, for the Type-B model, it is seen from Fig.~\ref{f3} that the CDF-\uppercase\expandafter{\romannumeral2} preferred region that explains the $W$-boson mass excess is entirely ruled out by the global fits of EW precision observables.

\section{Conclusion and Discussion}\label{s4}
In order to explain the excess of the $W$-boson mass recently observed by the CDF-\uppercase\expandafter{\romannumeral2} Collaboration, we have studied two promising candidate scenarios both involving high-dimensional $SU(2)_L$ scalar multiplets. %  detailed and concrete study to explain this anomaly in terms of high-dimensional $SU(2)_{L}$ scalar multiplets, considering the tree-level and one-loop level effects respectively. 
The first scenario is to introduce an odd-dimensional scalar multiplet with $Y=0$ so that the $W$ mass correction from its nonzero scalar VEV can explain the CDF-\uppercase\expandafter{\romannumeral 2} excess at the tree level. One salient feature of this mechanism is that the scalar VEV 
does not contribute to the mass of $Z$ boson, so that it can easily evade the associated strong constraint from $m_Z$. By estimating the experimental bound on the $\rho$ parameter through its relation to the $T$ parameter, we find that the regions permitted by the $\rho$ parameter is enlarged by about one order of magnitude, and thus the tree-level scalar multiplet scenario is available to explain the CDF-\uppercase\expandafter{\romannumeral2} $W$ mass anomaly even though the allowed parameter regions is strongly reduced.
% we have considered the scalar multiplet solution at the tree level, the excess of the $W$-boson mass can be explained by introducing an odd-dimensional $Y=0$ scalar multiplet with a non-zero vacuum expectation value that leaves the mass of the $Z$-boson unaffected. However, this scenario is constrained by the EW precise measurement of the $\rho$ parameter, rendering the problem unsolvable at the tree level alone. 
Then, we turn to the second scenario in which the CDF-\uppercase\expandafter{\romannumeral 2} anomaly could be possibly solved by one-loop effects generated by the additional scalar multiplet. We take a scalar octuplet as a concrete example to investigate this possibility. By making use of the octuplet scalars' contributions to the oblique parameters $S$ and $T$, it is found that the CDF-\uppercase\expandafter{\romannumeral2} measured $W$-boson mass can be easily obtained. We also take into account the constraints from the EW global fits and the perturbativity limit in this model. 
% one-loop , after considering the one-loop effects, the problem becomes tractable. The newly introduced scalar multiplet could correct the mass of the $W$-boson by contributing to the oblique parameters $S$ and $T$, resulting in a $W$-boson mass consistent with the measurement of the CDF Collaboration. We investigate this case phenomenologically by introducing a complex scalar nonet and taking into account constraints from EW global fits and perturbation limit. 
For the $Y=7/2$ case, the current data from the EW precision measurements has excluded the Type-B model in which the lightest particle is the electrically neutral one. % is completely excluded within the parameter space of $M_L$ from 500 to 4000 GeV by the EW global fit. Conversely,
In contrast, for the Type-A model in which the lightest scalar is the most charged one, it is found that there is a substantial parameter region explaining the CDF-\uppercase\expandafter{\romannumeral2} $m_W$ anomaly while still allowed by the experimental and theoretical constraints. In the latter case, the lightest scalar mass is found to lie in the range from 1800 to 5000 GeV.% the Type-A model with $M_L=M_C$ can survive within the range of $M_L$ from 1800 to 4000 GeV as well as fulfill the perturbation requirement. The parameter space below 1800 GeV is excluded by the electroweak global fit.

In this letter, we have exclusively investigated two representative cases, namely the high-multiplet scenarios with tree-level corrections and one-loop effects. In more generic cases, both mechanisms should work to solve the $W$-boson mass anomaly like a model added by a scalar triplet with $Y=0$~\cite{Cheng:2022hbo,FileviezPerez:2022lxp,Song:2022jns}. 
% Besides these two extreme cases, our results indicate that for an odd-dimensional scalar multiplet with $Y=0$, the anomalous $W$-boson mass can be explained by simultaneously considering the effects of both tree-level (i.e., introducing a sufficiently small vacuum expectation value) and one-loop corrections. 
Finally, we would like to mention that, for the available Type-A octuplet model with $Y=7/2$ and a vanishing scalar VEV, due to the accidental $Z_2$ symmetry, the lightest but also most charged particle $\xi^{\pm 7}$  should be nearly stable. %which means they are detector stable. 
This particle together with its less charged but heavier partners in the octuplet can be pair produced via the Drell-Yan processes. By introducing an additional high-dimensional effective operators such as ones in Ref.~\cite{Wu:2022uwk}, the $Z_2$ symmetry is explicitly broken and the lightest particle $\xi^{\pm 7}$ can decay into multiple-$W$ or multiple-lepton final states. However, owe to the suppression from the high dimension of operators as well as the multiplicity of produced particles in the final states, a simple estimate shows that $\xi^{\pm 7}$ should be long-lived with its lifetime of ${\cal O}$(s). This implies that $\xi^{\pm 7}$ at the LHC should be stable and only leaves charged tracks in the detectors~\cite{ATLAS:2023dnm,CMS:2016kce}. Currently, the most recent search at the LHC for such a highly charged particle has been performed in Ref.~\cite{Altakach:2022hgn}, which gives the lowest bound on its mass to be around 1700~GeV. For other charged counterparts such as $\xi^{\pm i}$ with $i\leq 6$, they would decay in a cascade into $\xi^{\pm 7}$ by emitting several $W$-bosons, giving a signature of multiple $W$'s plus charged tracks, which potentially serves as a key signal. The detailed investigation of these LHC signatures is well beyond the scope of our work, so that we leave them for future studies.   

%  could result in the multi-$W$ or multi-lepton signatures, which potentially serve as key signals for the present LHC and near-future collider searches. Currently,  and they are still allowed by the current LHC constraint when their mass is above 1700 GeV \cite{Altakach:2022hgn}.}

\section*{Acknowledgments}

This work is supported in part by the National Key Research and Development Program of China (Grant No.~2021YFC2203003 and No.~2020YFC2201501 )
and the National Natural Science Foundation of China (NSFC) (Grant No. 12005254 and No. 12147103).

%\begin{thebibliography}{0
\bibliography{mini-wmass}

%\end{thebibliography}

\end{document}